\documentclass[12pt]{iopart}
\usepackage{multirow,eurosym,amssymb,amsfonts,setspace,graphicx,bm,float,amssymb}

\usepackage{color}

\newcommand{\bra}[1]{\langle #1|}
\newcommand{\ket}[1]{|#1\rangle}

\def\be{\begin{equation}}
\def\ee{\end{equation}}

\def\bsplit{\begin{split}}
\def\nsplit{\end{split}}

\begin{document}
\title{Deterministic amplification of Schr\"odinger cat states in
  circuit quantum electrodynamics}

\date{\today}

\author{Jaewoo Joo} \address{Quantum Information Science, School
  of Physics and Astronomy, University of Leeds, Leeds LS2 9JT, United Kingdom}
\address{Advanced Technology Institute and Department of Physics,
  University of Surrey, Guildford, GU2 7XH, United Kingdom}

\author{Matthew Elliott} \address{Advanced Technology Institute
  and Department of Physics, University of Surrey, Guildford, GU2 7XH,
  United Kingdom}

\author{Daniel K L Oi} \address{SUPA Department of Physics,
  University of Strathclyde, Glasgow, G4 0NG, United Kingdom}

\author{Eran Ginossar} \address{Advanced Technology Institute and
  Department of Physics, University of Surrey, Guildford, GU2 7XH,
  United Kingdom}

\author{Timothy P Spiller} \address{York Centre for Quantum
  Technologies, Department of Physics, University of York, York YO10
  5DD, United Kingdom}

\begin{abstract} 
We propose a dynamical scheme for deterministically amplifying photonic Schr\"odinger cat states, based on a set of optimal state-transfer steps. Perfect deterministic amplification of arbitrary coherent states is prohibited by quantum mechanics but determinism can be achieved by sacrificing either fidelity or amplification factor. Our protocol is designed for strongly coupled circuit-quantum electrodynamics and utilises artificial atomic states and external microwave driving fields. In principle, high-fidelity amplification is possible with this protocol, while displaying a tradeoff between amplification and fidelity. We compare analytical results with full simulations of the open Jaynes-Cummings model with realistic device parameters compatible with the state of the art superconducting circuits. Amplification with a fidelity of 0.9 can be achieved for moderate size Schr\"odinger cat states in the presence of cavity and atomic-level decoherence. This amplification tool can be applied to practical quantum information processing with nonclassical continuous-variable states.
\end{abstract}


\maketitle 
\section{Introduction}
Quantum physics does not allow perfect deterministic amplification of unknown quantum states because additional quantum noise is inevitably introduced by the amplification process \cite{RMP}. The most commonly studied methods of high fidelity amplification of coherent states (i.e. $\ket{\alpha} \rightarrow \ket{G \alpha}$ for $G>1$) are based on probabilistic addition and subtraction of single photons \cite{add_subtract}. The fidelity and amplification factor $G$ of these processes vary differently with input amplitude $\alpha$, depending on the amplification operator that is implemented, for example $\hat{a} \hat{a}^{\dag}$ or $(\hat{a}^{\dag})^{2}$ \cite{NoiselessAmp_exp, OptAmp01, Jeong_NatPhys}. Such schemes are always restricted by the tradeoff between amplification factor and fidelity, as perfect amplification is forbidden by the no cloning theorem \cite{NoiselessAmp_exp}.

Superpositions of two large coherent states with opposite phases, called Schr\"odinger cat states (SCSs) \cite{Schro01,Peter05}, have great potential to open up new avenues for quantum technology, including continuous-variable (CV) quantum communication \cite{Polzik10}, fault-tolerant quantum computation \cite{Jeong02, Ralph03, cat_QC_threshold}, CV teleportation \cite{teleamp}, and quantum metrology \cite{Sanders12, JooPRL11}. There is therefore particular interest in deterministic amplification schemes for these states, in addition to studying fundamental aspects of amplification. If moderate sized SCSs - large enough that the coherent states have little overlap, but small enough to prevent excessive decoherence by photon loss - can be produced and stablised, then fault-tolerant CV quantum computing is possible using linear optics only. It is known that two identical SCSs can deterministically produce a larger SCS \cite{Andersen, HJ_original}, while several high-fidelity probabilistic methods of amplifying SCSs have recently been developed in quantum optics \cite{Grangier, teleamp}.

The recent, rapid development of superconducting circuit technology has the potential to provide a new platform for scalable quantum systems. The Josephson junction non-linearity allows the realization of superconducting artificial atoms (qubits) which can be strongly coupled to 3D cavities containing nonclassical microwave states. Sufficiently large SCSs for applications in quantum information ($\alpha \approx 2$) \cite{Yale_arxiv, Yale_big_cat} and generalized Fock states \cite{Hofheinz09} have recently been generated in microwave cavities with the assistance of superconducting qubits. Moreover, enhanced stabilisation of SCSs in a cavity has recently been reported using a specially designed lossy environment \cite{Yale_arxiv, 2photonloss, Footnote2}, with the aim of producing robust quantum memory~\cite{QEC_Yale}. Thus, amplification of SCSs would benefit a wide range of hybrid-state quantum technologies, and enable a new type of quantum computation within the framework of circuit-quantum electrodynamics (circuit-QED) \cite{NJP_Yale}. 
 
In this paper, we propose a scheme for amplifying an SCS in superconducting circuits. The key benefits of this atom-assisted method are that (A) it is deterministic: atomic excited states can repeatedly be prepared by controlled microwave pulses and the core of the amplification operation is performed as a unitary generalised $\sigma_x$ operation in the dressed representation of atomic and photonic states and (B) it does not require any specific loss engineering. Our scheme is inspired by the fact that applying the two-photon shift operation \cite{Peter05}, $(\hat{E}^\dag)^2:\ket{m}\longrightarrow\ket{m+2}~\forall m$, one or more times to an even/odd SCS preserves the even/odd distribution of number-state amplitudes. We analyse and simulate the operation $\hat{E}^{\dag}$ and $(\hat{E}^{\dag})^2$ acting on a fixed even SCS ($\alpha=1.5$) in a cavity coupled to a qubit in the presence of decoherence~\cite{BishopPhD}. For this scheme, we find fast microwave pulse controls which perform all the state-transfers required with high fidelity and within the decoherence time of realistic circuit-QED systems based on transmons and high-$Q$ cavities.

\section{Theoretical $(\hat{E}^{\dag})^{2}$ amplification of SCSs}
We generalize the notion of amplification to the case where an initial even/odd SCS,
\begin{equation}
|SC^{\pm}_{\alpha} \rangle = {\cal N}^{\pm}_{\alpha} (|\alpha \rangle \pm |{\rm -} \alpha \rangle),
\label{Back_01} 
\end{equation}
for some normalisation  ${\cal N}^{\pm}_{\alpha}$, is transformed by an operation $\hat{A}$ into a state $\hat{A}\ket{SC_{\alpha}^{\pm}}$, which approximates a larger SCS,
\begin{equation}
\ket{SC^{\pm}_{\alpha'}} \approx \hat{A}\ket{SC_{\alpha}^{\pm}} = \sum^{\infty}_{k=0} c_{k} \ket{k+b} \langle k \ket{SC^{\pm}_{\alpha}},
\label{Back_00}
\end{equation}
with $\alpha'> \alpha$ and $b>0$.  Due to destructive interference between $\ket{\alpha}$ and $\ket{-\alpha}$, even SCSs have only even photon numbers while odd SCSs have only odd photon numbers. Note that the amplitudes $c_k$ are determined by the amplification operator $\hat{A}$.

\begin{figure}[t]
\includegraphics[width=7.35cm]{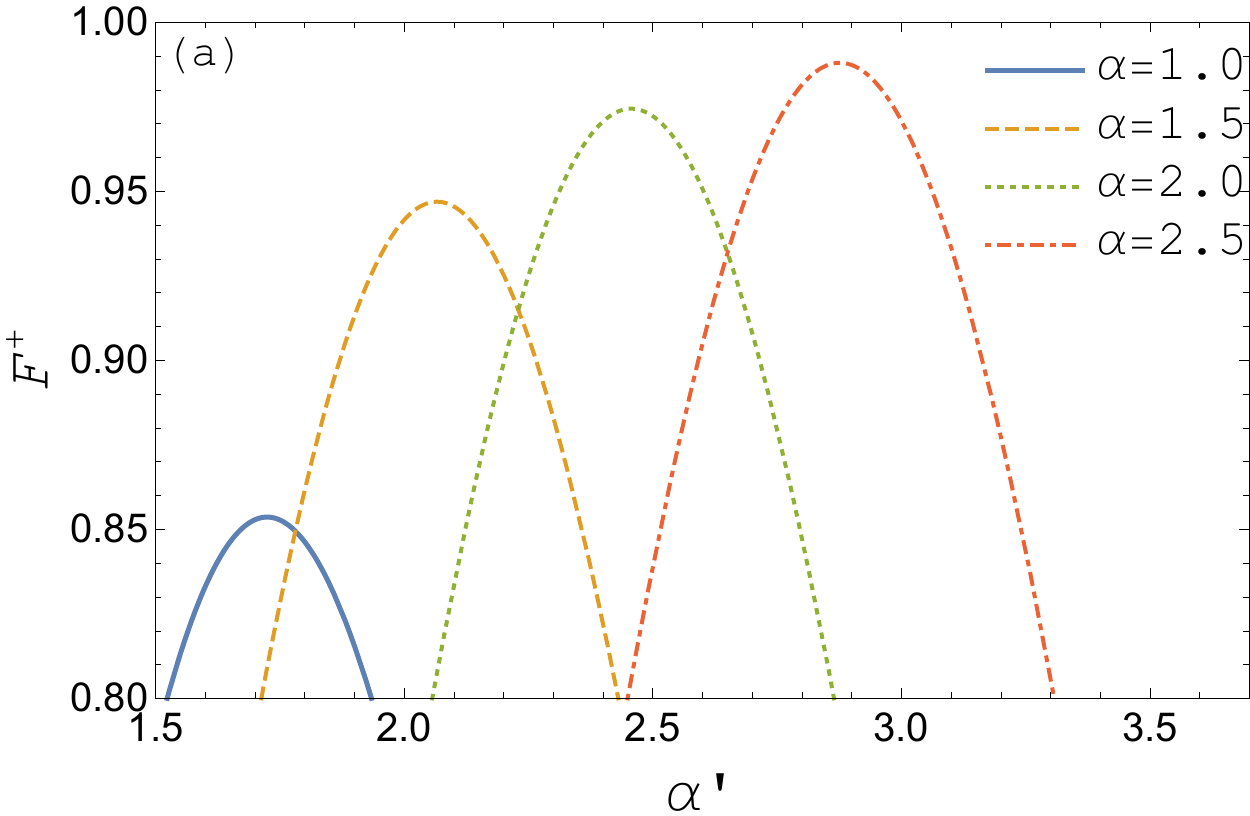}
\includegraphics[width=7.35cm]{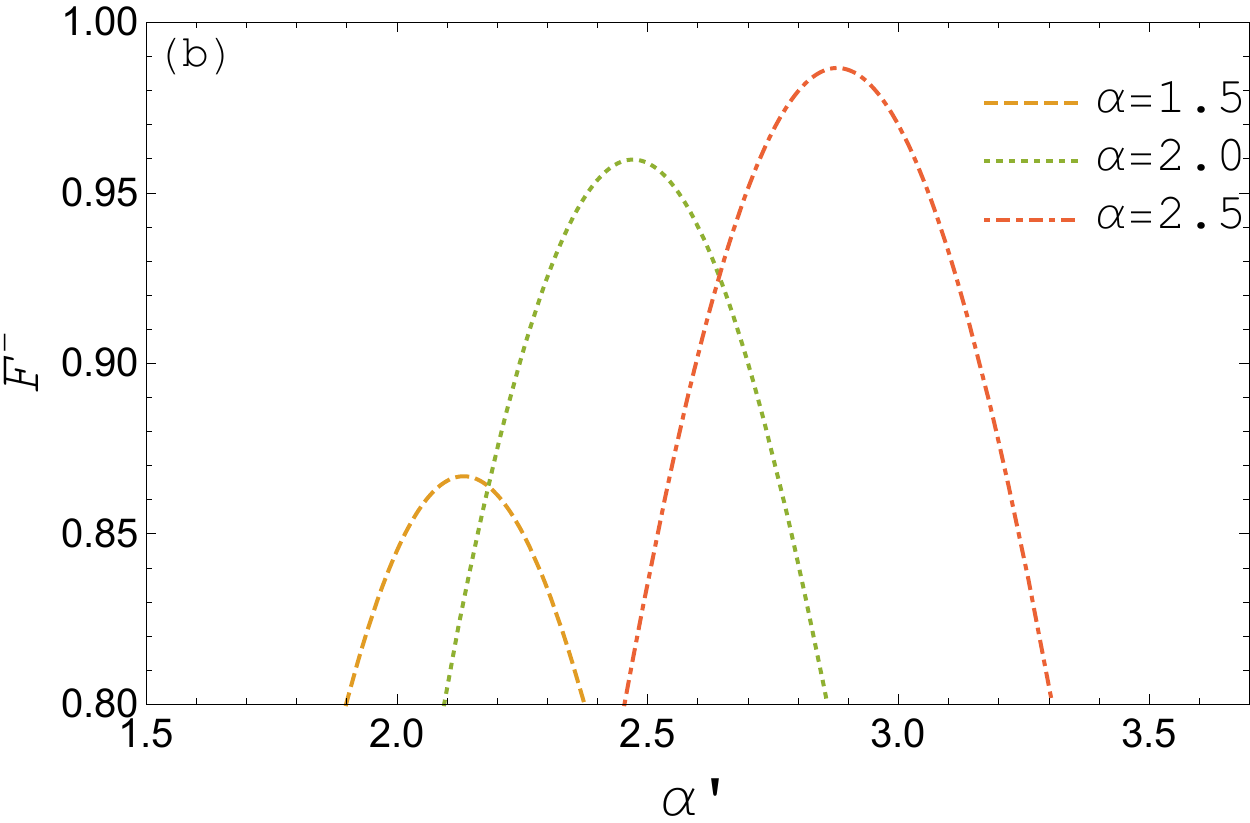}
\caption{Fidelities ${\cal F}^\pm$ between $(\hat{E}^\dag)^2|SC^\pm_\alpha \rangle$ and amplified state $|SC^{\pm}_{\alpha'} \rangle$ for $\alpha=1.0,1.5,2.0,2.5$. 
The maximum fidelities approach 1 for large $\alpha$ while the amplification rate defined in Eq.~(\ref{G_rate}) also goes to unity, demonstrating the fundamental tradeoff between the fidelity and amplification rate. For small $\alpha$, $(\hat{E}^{\dag})^2$ works better for even SCSs, but this difference between even and odd SCSs disappears for $\alpha > 1.5$ because $| \alpha \rangle$ is sufficiently orthogonal to $|{\rm-} \alpha \rangle$.}
\label{fig:02}
\end{figure}

If we choose the amplification operator to be the two-photon shift operator applied $l$ times,
\begin{equation} 
\hat{A} =  \left( \hat{E}^{\dag} \right)^{2l}=
\sum_{m=0}^{\infty} |m+2l\rangle\langle m|, \label{GeneralEdag01}
\end{equation}
the Fock state amplitude distribution is simply shifted and the normalisation of the outcome state is preserved. It can therefore be performed deterministically in principle \cite{Footnote01}. Fig. \ref{fig:02} shows the results of applying $(\hat{E}^{\dag})^2$ to both even and odd SCSs and calculating the overlap of an amplified SCS $(\hat{E}^{\dag})^2 \ket{SC^{\pm}_{\alpha}}$ with a target SCS $ \ket{SC^{\pm}_{\alpha'}}$, where the fidelities are
\begin{equation}
{\cal F}^{\pm} = \left|\langle SC^{\pm}_{\alpha'}| (\hat{E}^{\dag})^2 \ket{SC^{\pm}_{\alpha}} \right|^2.\label{Back_07} 
\end{equation}
We quantify the amplification by the value $G$, defined by $\alpha'=G\alpha$ which maximises $\mathcal{F}^\pm$, giving the closest SCS to $ (\hat{E}^{\dag})^2 \ket{SC_{\alpha}^\pm}$,
\begin{equation}
G=\arg \max_{G'} \left|\langle SC^{\pm}_{G' \alpha}| (\hat{E}^{\dag})^2
    \ket{SC^{\pm}_{\alpha}} \right|^2.
\label{G_rate}
\end{equation}

In general, the maximum fidelity ${\cal F}^{\pm}_{max}$ approaches 1 for large $\alpha$ but $G$ also tends to 1, indicating minimal amplification of very large SCSs, but stabilisation of the input SCS persists. We show fidelities between $(\hat{E}^\dag)^2|SC^\pm_\alpha \rangle$ and ideal amplified state $|SC^{\pm}_{G \alpha} \rangle$ for $\alpha=1.0,1.5,2.0,2.5$. The ${\cal F}^{+}_{max}$ are 0.854, 0.947, 0.974, 0.988, corresponding to $G \approx 1.725, 1.377, 1.229, 1.151$, while the ${\cal F}^{-}_{max}$ are 0.681, 0.866, 0.960, 0.987 with $G \approx 1.902, 1.422, 1.235, 1.151$. Interestingly, for $\alpha < 1.5$, $(\hat{E}^{\dag})^2$ works better for even SCSs because $\ket{SC^{-}_{\alpha\approx 0}}$ is approximately a one-photon Fock state. When amplified, this a mapped to a three-photon Fock state, which is very different to any odd SCS and we find that ${\cal F}^{-}_{max}<0.8$ for $\alpha = 1.0$. This behaviour disappears for $\alpha \ge 1.5$ because $| \alpha \rangle$ is sufficiently orthogonal to $|{\rm-} \alpha \rangle$. Thus, we will focus on how to implement the amplification procedure for an SCS with $\alpha=1.5$, in the range of interest for CV quantum information processing.

\begin{figure}[t]
\centering
\includegraphics[width=11cm]{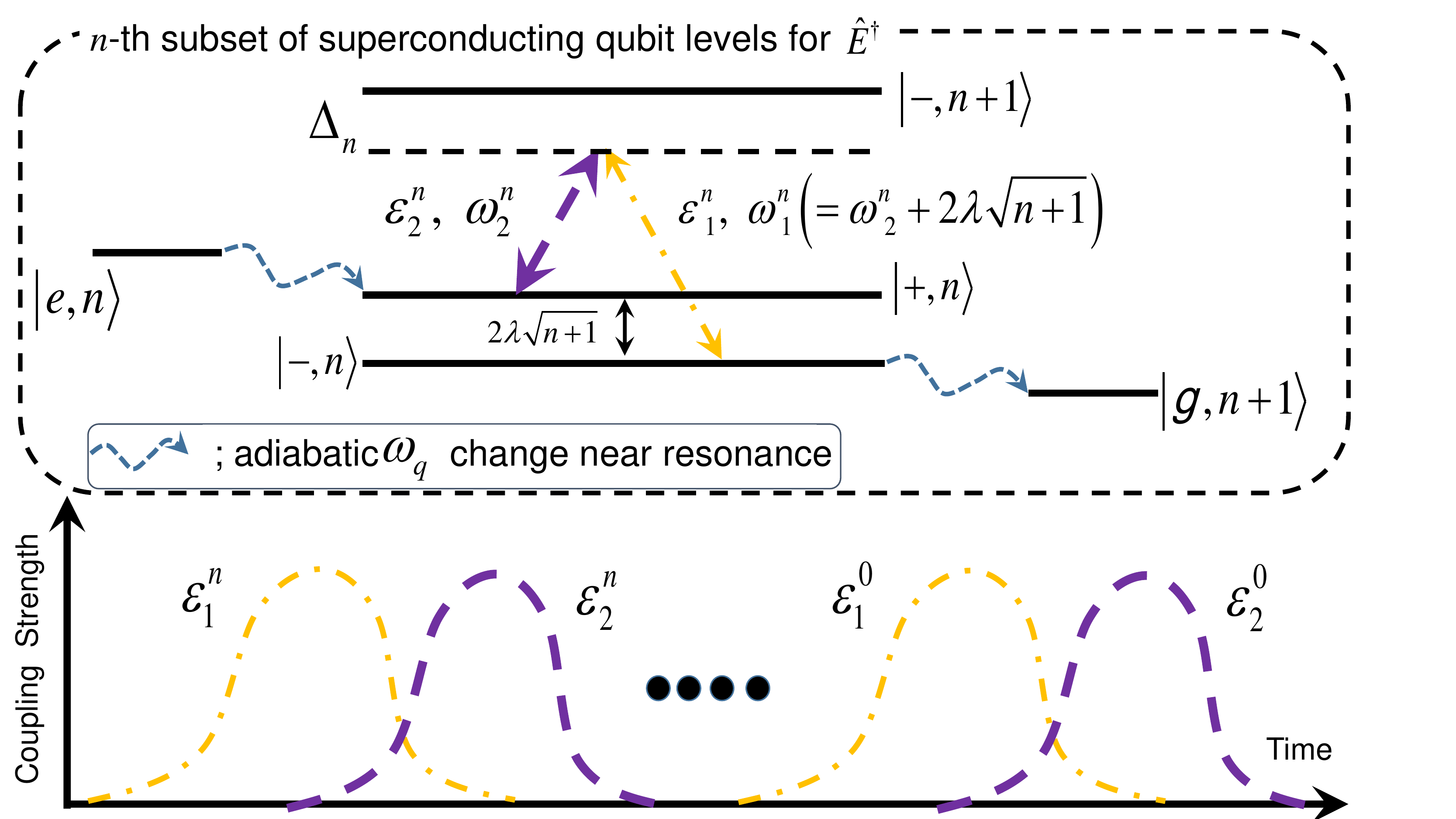}
\caption{STIRAP-type pulse sequence to realise $\hat{E}^{\dag}$ for n-th number state $\ket{n}$. (Top) An adiabatic sweep of the qubit frequency $\omega_{q}$ into resonance with the cavity $\omega_{r}$ transforms the initial state $\ket{e,n}$ into the dressed state $\ket{+,n}$. Next a microwave field is first applied to the $\ket{-,n} \leftrightarrow \ket{-,n+1}$ transition with time dependent Gaussian amplitude $\epsilon^{n}_1(t)$ and frequency $\omega^n_1$ (yellow dot-dashed line), followed by another field driving the second transition ($\ket{+,n} \leftrightarrow \ket{-,n+1}$) with envelope $\epsilon^{n}_2(t)$ and frequency $\omega^n_2$ (purple dashed line). For an SCS, the $\ket{-,n+1}$ state is unpopulated hence does not participate in the dynamics. The microwave frequencies are detuned $\Delta_n$ from the $\ket{-,n+1}$ state but satisfy the two-photon transition condition, $\omega^n_1-\omega^n_2 = 2\lambda \sqrt{n+1}$. After the pulse sequence, a further adiabatic sweep of the transmon frequency back out of resonance disentangles the atom from the cavity, resulting in the state $\ket{g,n+1}$. The action on the cavity state is $\ket{n} \rightarrow \hat{E}^{\dag} \ket{n}=\ket{n+1}$. (Bottom) If  the input state is an even SCS given by $\sum c_n \ket{n}$, a set of pulses $\{ \epsilon^{n}_1(t), \epsilon^{n}_2(t) \}$ acting on each number state produces the outcome state $\sum c_n \ket{n+1}$.}
\label{fig:levels-STIRAP}
\end{figure}

\section{Implementation in circuit-QED}
Circuit-QED provides an ideal regime for amplification of SCSs, due to the large nonlinearities and strong coupling that can be achieved. We first outline our scheme for deterministically performing a single $\hat{E}^{\dag}$ operation on a cavity field, with further details of the implementation in the following sections. This operation can be applied twice to achieve $(\hat{E}^{\dag})^2$, and therefore amplification of SCSs. The protocol, shown in Fig.~\ref{fig:levels-STIRAP}, is as follows: (1) an SCS $\ket{SC^{\pm}_{\alpha}} =\sum_{n=0}^{\infty} c_{n} \ket{n}$ is initially prepared in the cavity, with the qubit in $\ket{e}$, where the $c_{n}$ vanish for odd (even) $n$ for even (odd) SCSs.(2) An adiabatic sweep is used to bring the qubit frequency $\omega_q$ from far off-resonance to the resonator frequency $\omega_r$, where the eigenstates of the system are dressed qubit-cavity states \cite{BishopPhD}. This slowly transfers the bare system into a superposition of dressed states $\sum_{n=0}^{\infty} c_{n} \ket{+,n}$. (3) A state-transfer scheme adapted from the original idea of stimulated Raman adiabatic passage (STIRAP) in cavity-QED \cite{Kuhn1999, Zoller_Kimble95, ZouGuo_PLA06, OJP2013} is performed. Instead of using a bare atomic $\Lambda$-level configuration, we use a set of $\Lambda$-type systems in the dressed Jaynes-Cummings (JC) model, with dynamical control provided by varying local fluxes~\cite{supercon_implement, Eran_PRA10}. Pairs of overlapping Gaussian microwave pulses are applied to the effective three-level systems $\{\ket{+,n},\ket{-,n},\ket{-,n+1}\}$ to transfer the populations into to state $\sum_{n=0}^{\infty} c_{n} \ket{-,n}$, via the $\ket{-,n+1}$ states (4) The first step is reversed, sweeping $\omega_q$ away from $\omega_r$ to disentangle the final cavity state from the qubit, leaving it in the state $\sum_{n=0}^{\infty} c_{n} \ket{n+1}\approx \ket{SC^{\mp}_{\alpha'}}$ which has $\alpha'>\alpha$ and opposite parity.

To repeat the operation, the qubit must be reset from $\ket{g}$ to $\ket{e}$ by a further microwave pulse, but is now sufficiently far detuned that the cavity state is not affected. Finally, at the end of the second $\hat{E}^{\dag}$ operation, we consider the Selective Number-dependent Arbitrary Phase (SNAP) gate to correct relative phases between Fock states acquired during the operation. This technique has been already demonstrated in a protocol to minimize phase distortions due to self-Kerr interactions \cite{SNAPgate}.

\subsection{Adiabatic sweep of qubit frequency $\omega_{q}$}
\begin{figure}[t]
\center
\includegraphics[width=11cm]{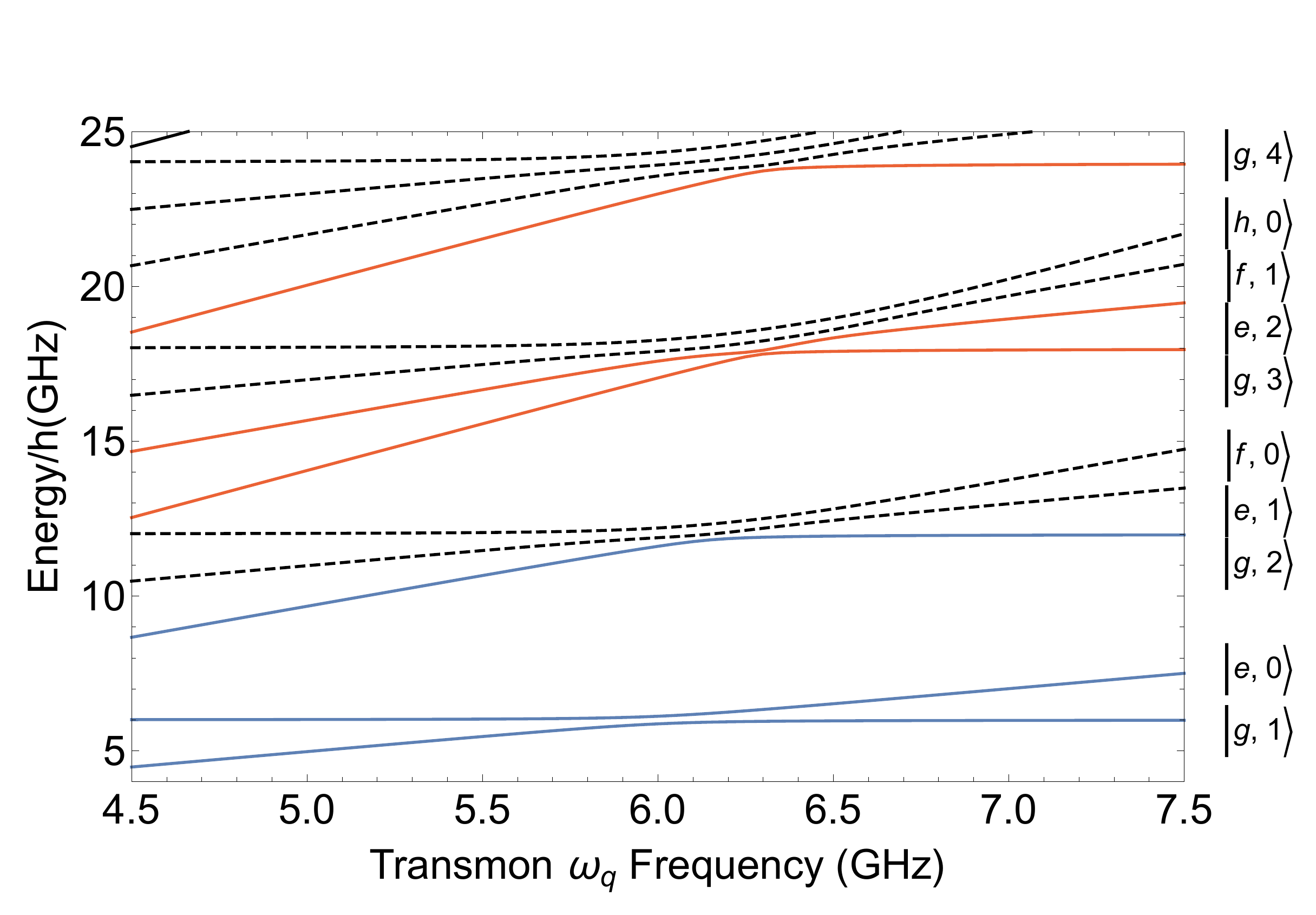} 
\caption{Energy level structure of transmon coupled to a cavity with $\omega_r /2\pi = 6$GHz and $\lambda /2\pi = 0.1$GHz \cite{supercon_implement}. Solid lines indicate two sets of $\Lambda$-type dressed levels $\{\ket{+,n},\ket{-,n},\ket{-,n+1}\}$ suitable for state-transfers and dashed lines are other eigenstates of the Hamiltonian in Eq.~(\ref{trans_Ham})~\cite{BishopPhD,JohnsonPhD}. The labels on the right hand side are the product states that approximate the eigenstates for large positive detunings ($(\omega_q-\omega_r)/ \lambda \gg 1$).}
\label{fig:transmon_levels}
\end{figure}

We model a transmon qubit coupled to a cavity by a generalised JC Hamiltonian
\begin{equation}
\label{trans_Ham}
\hat{H}^{t} = \omega_{r} \hat{a}^{\dag} \hat{a} + \sum_{j} {\omega_{qj} \over 2}
\ket{j}\bra{j} + \sum_{j,k} \lambda_{j,k} (\hat{a}^{\dag}  \ket{j}\bra{k}
+ \hat{a} \ket{k}\bra{j}),
\end{equation}
for transmon energy levels $j,k= \{g, e, f, h, ...\}$ and transmon-cavity couplings $\lambda_{j,k}$. As shown in Fig.~\ref{fig:transmon_levels}, when the transmon frequencies are far from resonance with the cavity, the bare states are given by $\ket{j,n}$ with transmon state $j$ and photon number $n$, while they become dressed states near resonance. Considering only two transmon levels, $\{\ket{g},\ket{e}\}$, the eigenstates are
\begin{eqnarray}
\label{Dressed_state}
\ket{+,n} &=& \cos \theta_n \ket{e,n} +\sin \theta_n \ket{g,n+1}, \\
\ket{-,n} &=& -\sin \theta_n \ket{e,n} +\cos \theta_n \ket{g,n+1}, 
\end{eqnarray}
where  $\omega_{q}=\omega_{qg}$ is the $\ket{g}\leftrightarrow\ket{e}$ transition frequency, $\lambda=\lambda_{g,e}$ is the qubit-cavity coupling, and $2 \theta_n = \tan^{-1} (2 \lambda \sqrt{n+1}/\delta)$, with $\delta = \omega_q - \omega_r$.  Note that $\ket{+,n} \approx \ket{e,n}$ and $\ket{-,n} \approx \ket{g,n+1}$ for large $\delta$, so if we start in $\ket{e,n}$ far from resonance, the state adiabatically becomes $\ket{+,n}$ as $\omega_r$ approaches resonance $\delta \approx 0$. This process requires the use of flux-tunable qubits. In realistic devices with multiple transmon levels, this sweeping must be performed slowly enough to prevent leakage of population to higher levels.

\subsection{Protocol for state-transfer on even SCSs in circuit-QED}
\label{III-B}
The key element of our $\hat{E}^{\dag}$ operation is an efficient state transfer from $\ket{+,n}$ to $\ket{-,n}$ which is performed on individual number states using $\Lambda$-type sets of levels. Once the initial, dressed state is prepared a microwave field is first applied between $\ket{-,n}$ and $\ket{-,n+1}$ with time dependent amplitude $\epsilon^{n}_1 (t) =|\epsilon^{n}_{1}| \exp\left[ -{(t-\tau)^2/ T^2} \right]$ and frequency $\omega^n_1$, followed by another field driving the $\ket{+,n} \leftrightarrow \ket{-,n+1}$ transition ($\epsilon^{n}_2(t) =|\epsilon^{n}_{2}| \exp\left[ -{(t+\tau)^2/T^2} \right]$, $\omega^n_2$). Both drives are detuned by $\Delta_n$ from their respective transitions, while still satisfying the two photon condition $\omega^n_1-\omega^n_2 = 2\lambda \sqrt{n+1}$, which ensures that the intermediary $\ket{-,n+1}$ state is not populated. The pulses have non-zero overlap, determined by the temporal offset $\tau$. For efficient transfer of $\ket{+,n}\rightarrow\ket{-,n+1}$, we require $\tau > (\sqrt{2} -1)T$ and $|\epsilon| T \gg 10$~\cite{STIRAP_Supercond_01}. 

\begin{figure}[b]
\centering
\includegraphics[width=\columnwidth,clip=true,trim=0cm 2cm 0cm 1.5cm]{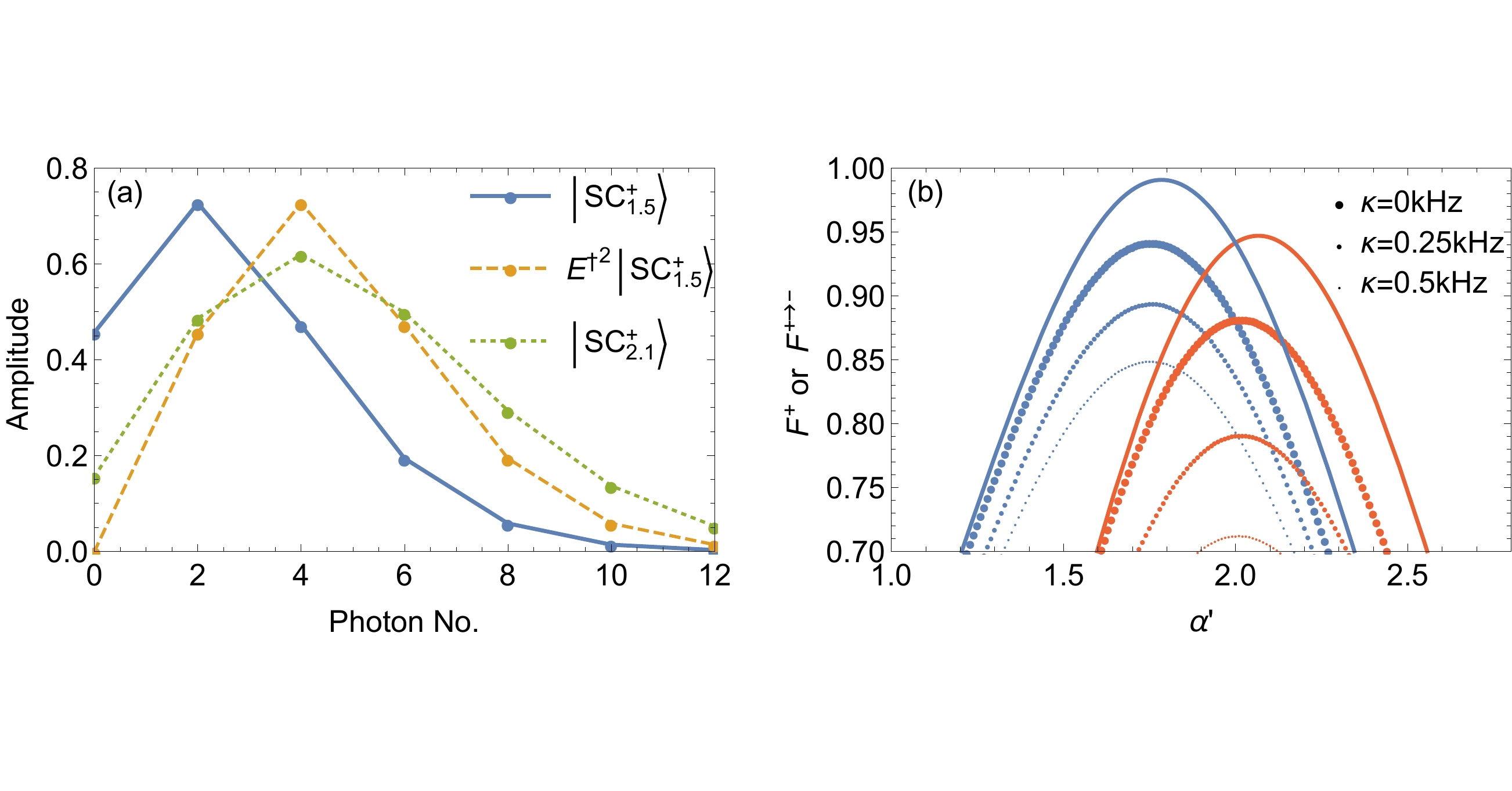}
\caption{(a) Photon number amplitudes for states $\ket{SC^{+}_{1.5}}$,  $(\hat{E}^{\dag})^2\ket{SC^+_{1.5}}$ and $\ket{SC^+_{2.1}}$. We see that the population of the 8-photon Fock state is lower than 1\% of the population for $\ket{SC^{+}_{1.5}}$ and that the four sets of state-transfers cover enough states for amplifying these states. (b) Fidelities ${\cal F}^{+\rightarrow-}$ (blue) and ${\cal F}^{+}$ (red) achieved by applying $\hat{E}^{\dag}$ and $(\hat{E}^{\dag})^2$ respectively to $|SC^{+}_{1.5} \rangle$. The solid lines show the theoretical bounds, which are ${\cal F}^{+\rightarrow-}_{max}>$ 0.99 at $\alpha' \approx 1.78$ and ${\cal F}^{+}_{max} >$ 0.945 at $\alpha' \approx 2.1$. The dotted lines show the simulated performance with different values of decoherence ($\gamma_{-} =\gamma^{\phi}=10 \kappa$) using four sets of simultaneous state-transfer operations. An amplification factor of $G\approx 1.33$ is achieved for $(\hat{E}^{\dag})^2$.}
\label{fig:03}
\end{figure}

A pair of pulses is used to transfer each number state with significant population. For an even SCS, all odd-number states are unpopulated, so the $\Lambda$ systems are effectively independent of each other. This spacing of occupied and unoccupied states also prevents spectral crowding and the driving of unwanted transitions. The pulses are performed in the manifolds of dressed states $\{\ket{+,n},\ket{-,n},\ket{-,n+1}\}$, in order, from the $n$-th to 0-th manifold. After all pulse sets and disentanglement from the qubit, this leaves the final state $\sum c_n \ket{n+1}=\hat{E}^{\dag} \sum c_n \ket{n}$. We see that $\hat{E}^{\dag}$ flips the even SCSs to odd. Note that the analytical (theoretical) fidelity between $\ket{SC^{-}_{\alpha'}}$ and
$\hat{E}^{\dag}\ket{SC^{+}_{\alpha}}$ is given by
\begin{eqnarray}
{\cal F}^{+\rightarrow - } = \left|\langle SC^{-}_{\alpha'}\,| \hat{E}^{\dag} \ket{SC^{+}_{\alpha}} \right|^2.
\label{Back_06} 
\end{eqnarray}
Fig.~\ref{fig:03}(b) shows that the maximum fidelity of ${\cal F}^{+\rightarrow  -}$ of the theoretical bound is higher than 0.99 between $\hat{E}^{\dag} \ket{SC^{+}_{1.5}}$ and $\ket{SC^{-}_{1.78}}$.

The scheme discussed here is compatible with the existing protocol for creating SCSs in Ref. \cite{Yale_big_cat}, while conventional STIRAP and two-tone red sideband transitions have also been demonstrated in $\Lambda$-type superconducting systems \cite{STIRAP_Supercond_01, STIRAP_Masluk12,Leek09}.

\subsection{Protocol for $(\hat{E}^{\dag})^2$ and SNAP gates}
In contrast to the original cavity-QED proposal \cite{Zoller_Kimble95}, $\pi$ pulses can be used to reset the qubit state $\ket{g} \rightarrow \ket{e}$ directly without affecting the cavity state \cite{Yale_big_cat}, and hence $(\hat{E}^{\dag})^2$ can, in principle, be performed by repeating the protocol. The fidelity of the protocol is reduced by Kerr-type non-linearities in the dressed cavity, causing defects which accumulate over time. However, these distortions can be significantly remedied by using a corrective series of SNAP gates to compensate for the relative phases acquired by different number states \cite{SNAPgate}. As the procedure is limited by the decoherence time and these distortions, we examine both $\hat{E}^{\dag}$ and $(\hat{E}^{\dag})^2$ including qubit and cavity decoherence, along with corrections by SNAP gates.

\section{Simulation with decoherence} 
To examine the performance of the protocol, we numerically simulate a simplified driven JC Hamiltonian with two atomic levels ~\cite{Alsing1992,Blais04}
\begin{equation}
\label{JC_Ham}
\hspace{-2.5 cm}
\hat{H}^{tot} = \omega_{r} \left( \hat{a}^{\dag} \hat{a}+{ 1\over 2} \right) + {\omega_{q} \over 2} \hat{\sigma}^z + \lambda (\hat{a}^{\dag}  \hat{\sigma}^{-}
+ \hat{a} \hat{\sigma}^{+})  + \sum_{n} \sum_{j=1}^2 \epsilon^{n}_{j} (t) \left(e^{-i \omega^{n}_j t} \, \hat{a} + e^{i \omega^{n}_j t} \, \hat{a}^{\dag} \right),
\end{equation}
where $\omega_j^n$ are the frequencies of the microwave drives, and $\epsilon_j^n$ their amplitudes.  While the microwave driving terms couple all of the excitation subspaces of the undriven JC Hamiltonian, the Hamiltonian is only slightly perturbed for small $|\epsilon_j|$ and the pulse frequencies are far off-resonance from unwanted transitions. Thus, the majority of the evolution is confined to the respective $\{\ket{\pm,n}\}$ manifold. The bichromatic driving induces the transition $\ket{+,n}\rightarrow\ket{-,n}$ via quasi-adiabatic following even though the system is not at, or close to, an eigenstate of the instantaneous Hamiltonian for part of the pulse sequence. This exploits the topological properties of the dressed eigenenergy surfaces \cite{Amniat2004}. We briefly note that this procedure in the driven JC system has a different character to conventional STIRAP on a bare $\Lambda$-level atom with directly driven transitions and behaves reversibly due to the unitary nature of the evolution (see Fig.~\ref{STIRAP_reverse} in Appendix A).

\subsection{ $\hat{E}^{\dag}$ and $(\hat{E}^{\dag})^2$ with SNAP gates on $\ket{SC^{+}_{1.5}}$ }
We first simulate a single $\hat{E}^{\dag}$ operation acting on $\ket{SC^{+}_{1.5}}$. In order to perform $\hat{E}^{\dag}$ efficiently and practically, the minimum number of STIRAP-type sets can be decided by the plot of amplitudes of SCSs. Fig.~\ref{fig:03}(a) shows that $\ket{SC^{+}_{1.5}}$ has most of its population in four Fock states, $\{ \ket{0}, \ket{2}, \ket{4}, \ket{6}\}$, and therefore four sets of STIRAP-type pulses will cover enough population to achieve good amplification. To minimise the length of the procedure and hence to reduce decoherence to practical levels for $\hat{E}^{\dag}$, we perform all the transfers simultaneously, sharing a common first pulse. This produces almost identical fidelities to four independent state-transfer sets in the decoherence-free case, with large improvements when decoherence is included.

Our simulated system has $\lambda/2\pi =0.1$ GHz and $\omega_r /2\pi =6.0$ GHz. We start with the qubit 1 GHz detuned and perform the adiabatic sweep in 6.2 $\mu$s, which is sufficiently slow to prevent population transfer to unwanted levels. For the four sets of state transfers, we use a single $\omega_1$ which is shared between all transfers, adjusting $\Delta_i$ for each $\Lambda$-level system to find the appropriate value of $\omega_2^i$. For the the first $\hat{E}^{\dag}$ we use $\tau= 3.58 \mu s$ and $T=6.28 \mu s$, with other parameters given in Table \ref{tab:params}. With these parameters, the total state-transfer time is approximately 25 $\mu s$, which could be reduced further by using a larger transmon-cavity coupling strength. For the second $\hat{E}^{\dag}$, $\tau= 3.14 \mu s$ and $T=6.28 \mu s$. In the Fock basis, SNAP gates are given by $S_{snap} = \sum_{m=0} e^{i \Phi_{m}} \ket{m}\bra{m}$. The values of the phases are dependent on Fock states $\ket{m}$ and we choose $\{ \Phi_m |m=0,1,...,8 \}$ as given in Appendix B.

\begin{figure}[t]
\centering
\includegraphics[width=8cm,trim=0cm -0.5cm 0cm 0cm]{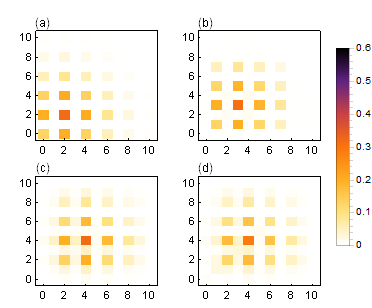}
\includegraphics[width=7.5cm,trim=0cm 0cm 0cm 0cm]{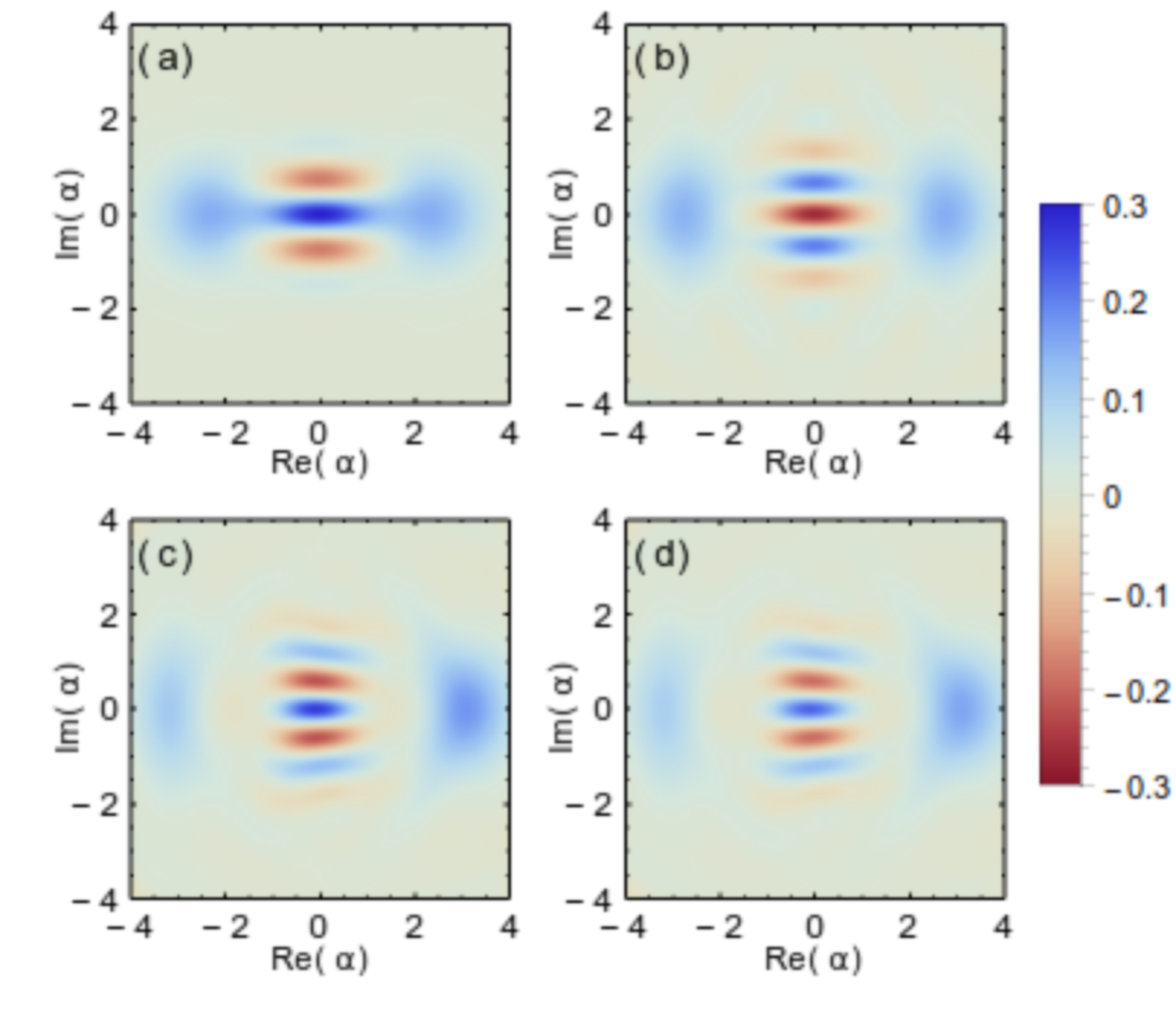}
\caption{Density matrix plots $c_{nm}$ (left) and Wigner functions (right). The initial state $\ket{SC^{+}_{1.5}}$ is plotted in (a), with numerical $\hat{E}^{\dag}\ket{SC^{+}_{1.5}} \approx \ket{SC^{-}_{1.8}}$ without decoherence shown in (b). Simulated two-photon shift with  $(\hat{E}^{\dag})^2$ is shown in (c) and, finally, (d) shows the final state $(\hat{E}^{\dag})^2\ket{SC^{+}_{1.5}} \approx \ket{SC^{+}_{2.1}}$ including decoherence of $\kappa/2\pi$= 0.25 kHz. On density matrix plots, the shifted blocks clearly indicate the $\hat{E}^{\dag}$ and $(\hat{E}^{\dag})^2$ are performed, with blurred regions caused by imperfect state-transfers and decoherence. In Wigner function plots, interference fringes can be clearly seen, with the central fringe changing color between even and odd SCSs, while some distortions are caused by imperfect transfers. Decoherence slightly reduces the color contrast of fringe patterns in (c) and (d).
\label{fig:055}}
\end{figure}

To model Markovian decoherence associated with cavity and qubit decoherence, we use a Lindblad master equation,
\begin{equation}
\dot{\rho} = i \left[ \hat{H}^{tot}, \rho \right] + \kappa {\cal D} [\hat{a}] + \gamma_- {\cal D} [\hat{\sigma}^-]  + {\gamma_\phi \over 2} {\cal D} [\hat{\sigma}^z]  ,
\label{master_01} 
\end{equation}
where ${\cal D} [\hat{b}] = \hat{b} \rho \hat{b}^{\dag} - { 1 \over 2} \{ \hat{b}^{\dag} \hat{b}, \rho\} $. 
We choose $\gamma_{-}/2\pi = \gamma^{\phi}/2\pi = 10 \kappa / 2\pi = 2.5, 5.0$kHz using realistic parameters from Ref.~\cite{overview}. The results are shown in Figs.~\ref{fig:03} and \ref{fig:055}. In Fig.~\ref{fig:03}(a), a comparison of the photon number amplitudes for $\ket{SC^{+}_{1.5}}$, $(\hat{E}^{\dag})^2\ket{SCS^+_{1.5}}$ and $\ket{SC^{+}_{2.1}}$ indicates the similarity of the amplified SCS and the desired target SCS. As shown in Figs.~\ref{fig:03}(b), $\hat{E}^{\dag}$ without decoherence achieves a maximum fidelity above 0.94, with the gap between the theoretical and no-decoherence cases caused by imperfections in the transfer method and partly due to a small population of higher dressed states over $\ket{+,8}$, which are not transferred. The dotted points show that decoherence almost linearly reduces the fidelity, with ${\cal F}^{+\rightarrow-}\approx 0.9$ and ${\cal F}^{+}\approx 0.8$ for $\kappa/2\pi$= 0.25 kHz.

\subsection{Evidence of performing $(\hat{E}^{\dag})^k$ on $\ket{SC^+_{1.5}}$ ($k=1,2$)}
It is straightforward to show the performance of the shift operation $(\hat{E}^{\dag})^k$ by looking at the density matrices of the initial and final states, because the components of the density matrix of SCSs are shifted by $k$ Fock-basis elements. For example, the density matrix of an initial even SCS is
\begin{eqnarray}
\rho_{int} &=& \ket{SC^+_{\alpha}} \bra{SC^+_{\alpha}}=\sum_{n,m=0}^{\infty} c_{nm} \ket{2n}\bra{2m}.
\end{eqnarray}
After the shift operation has been performed, the outcome state is given by
\begin{eqnarray}
\rho_{out} &=& (\hat{E}^{\dag})^k \ket{SC^+_{\alpha}} \bra{SC^+_{\alpha}} (\hat{E})^k =\sum_{n,m=0}^{\infty} c_{nm} \ket{2n+k}\bra{2m+k}. 
\end{eqnarray}

Plots of magnitude of density matrix elements (left side) and Wigner plots of the states (right side) are shown in Fig. \ref{fig:055}. Panels marked (a) show the initial SCS, which one can assume is prepared using the method of Ref.~\cite{Yale_big_cat}. In (b) the state after applying a single $\hat{E}^{\dag}$ followed by SNAP gates is shown. In (c), the $(\hat{E}^{\dag})^2$ operation is performed with a qubit flip after the first $\hat{E}^{\dag}$ and the SNAP gates after the second $\hat{E}^{\dag}$. Finally (d) includes decoherence with $\kappa/2\pi=0.25$ kHz  

We see that without decoherence the coefficients $c_{nm}$ are preserved but shifted. With decoherence, there is blurring of this effect as odd number states become populated. This type of quantum process tomography can be performed experimentally. In the Wigner functions, the interference fringes clearly switch between negative and positive values as the SCS switches between odd and even, with a central blue peak in (a), (c) and (d) while in (b) the central peak is red. In (d), we see that the decoherence slightly reduces the contrast of the fringe patterns but the state is clearly very similar to the no-decoherence case shown in (c). Using Ramsey interferometry, one can measure the qubit state-dependent phase shift of the cavity state, as explained in the Supplementary Material of Ref.~\cite{Yale_big_cat}, and therefore perform tomography on the state in the high-Q cavity via a low-Q cavity. Alternatively, a parity measurement can also be used to show the parity difference between the initial and final states \cite{6}.

\section{Summary and remarks } 
We have demonstrated a scheme for deterministic amplification of microwave SCSs using $\hat{E}^{\dag}$ and $(\hat{E}^{\dag})^2$ operations in circuit-QED. A STIRAP-type state transfer method provides the core of the $\hat{E}^{\dag}$ operation, which simply shifts the Fock state amplitude distribution of the initial SCS. The theoretical scheme was compared with the simulation of the Jaynes-Cummings model with three values of decoherence. The application of $\hat{E}^{\dag}$ amplifies the even SCS $\ket{SC^+_{1.5}}$ to the odd SCS $\ket{SC^-_{\approx 1.8}}$ with the fidelity 0.9 under realistic decoherence, whilst $(\hat{E}^{\dag})^2$ produces $\ket{SC^+_{\approx 2.0}}$ with fidelity 0.8. Due to the benefits of the superconducting system, this deterministic method might overcome the barrier to probabilistic amplification in optics-only methods for other applications in the future. Dissipation-egineering solutions provide a complementary scheme for continuously amplifying SCSs \cite{Yale_arxiv}, while our discretized scheme might be extendable to amplification of a bipartite (or multipartite) entangled SCSs without specially-designed lossy environments \cite{Sanders12, JooPRL11}. We note that a related experiment has been recently performed in coherent states in an ion-trap system \cite{iontrap}.

In CV quantum information processing using SCSs, $\hat{E}^{\dag}$ can be used as a bit-flip operation, by switching the state parity with minimal amplification for $\alpha > 2$, while $(\hat{E}^{\dag})^2$ can act as a stabilizer operation on SCSs. If one can perform either $\hat{E}^{\dag}$ or $(\hat{E}^{\dag})^2$ depending on the outcome of a parity measurement of the cavity state, it can be used for a discretized purification of SCSs. Taking advantage of well-separated lower energy levels, fluxonium or flux qubits could also be used in this scheme \cite{Fluxonium}. 

\section{ Acknowlegements} 
We would like to thank J. Dunningham for early contributions to this project and H. Jeong and B. Vlastakis for useful comments. EG acknowledges support from EPSRC (EP/L026082/1). DKLO acknowledges the Quantum Information Scotland (QUISCO) network.

\appendix
\section{Evidence of STIRAP-type operations}

Although this STIRAP-type operation behaves well enough for our desired state-transfer ($\ket{+,n} \rightarrow \ket{-,n}$), it cannot be fully explained by conventional STIRAP in a bare $\Lambda$ atomic system. In STIRAP, the state tranfer efficiency is strongly dependent on the overlap of the two pulse envelopes \cite{1}. In particular, efficient state-transfer only occurs for the counter-intuitive sequence of the two pulses ($\epsilon_{1}$ first and $\epsilon_{2}$ second). 

\begin{figure}[t]
\centering
\includegraphics[width=10cm]{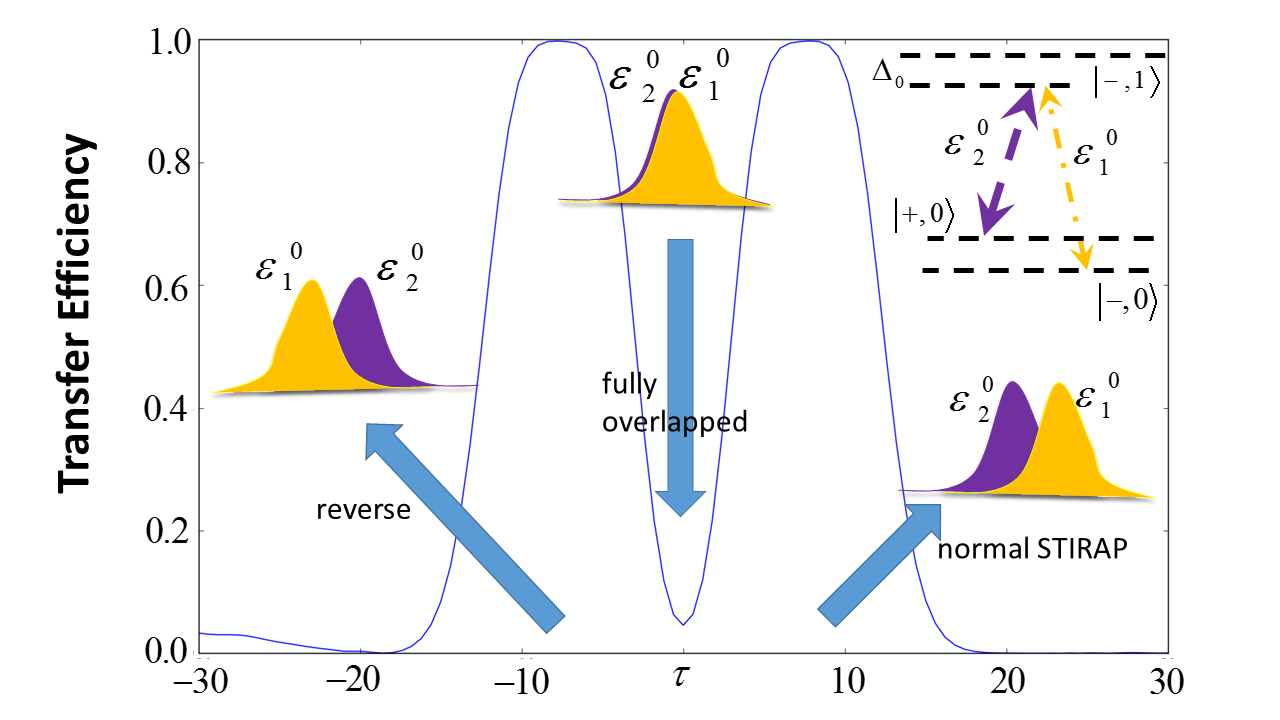}
\caption{Transfer efficiency of the STIRAP-type pulses between $\ket{+,0}$ and $\ket{-,0}$ as a function of the overlap between the two pulses (see the right top). It is the evidence that our state-transfer scheme is a generalized $\sigma_x$ operation between $\ket{+,n}$ and $\ket{-,n}$ for different $n$. \label{STIRAP_reverse}}
\end{figure}

We have examined the transfer efficiency of our scheme for the simplest transfer from $\ket{+,0}$ to $\ket{-,0}$ with detuning $\Delta_0$ in Fig.~\ref{STIRAP_reverse}. For positive $\tau$, the behaviour is similar to the normal STIRAP counter-intuitive pulse sequence, with transfer efficiency rapidly increasing as $\tau$ increases, nearly reaching 1 plateaing. The efficiency then drops with decreasing overlap area. However, our operation also shows excellent state-transfer for the reverse pulse sequence. It therefore behaves like a generalisation of the $\sigma_x$ operation to the cavity state.

In our parameter region, and without decoherence, the transfer efficiency is symmetric about $\tau=0$ (fully overlapped pulses). However, the transfer efficiency for reversed pulses is more sensitive to changes in $\Delta_0$ and the length of pulse envelopes. Oscillations are seen in the transfer efficiency, indicating that the process may not be `as adiabatic' as conventional STIRAP. These phenomena might be better understood in adiabatic Floquet theory \cite{2} and we believe they are caused by the existence of energy levels outside the $\Lambda$-system \cite{3}.

\section{Simulation parameters for SNAP gates and state-transfers}

\begin{table}[h]
\centering
\begin{tabular}{c|c|c|c|c|c}
 Transfer & $\epsilon_1 /2\pi$ (MHz)  & $\omega_1 /2\pi$ (GHz) & $\epsilon_2/2\pi$ (MHz) & $\omega_2/2\pi$ (GHz)& $\Delta/2\pi$ (MHz) \\
 \hline
$\ket{0}\rightarrow\ket{1}$ & 10 & 5.949 & 35 & 5.749 & 10\\
$\ket{2}\rightarrow\ket{3}$ & 10 & 5.949 & 38 & 5.603 & 24\\
$\ket{4}\rightarrow\ket{5}$ & 10 & 5.949 & 49 & 5.501 & 30\\
$\ket{6}\rightarrow\ket{7}$ & 10 & 5.949 & 70 & 5.419 & 33\\
\hline
$\ket{1}\rightarrow\ket{2}$ & 24 & 5.953 & 55 & 5.679 & 15\\
$\ket{3}\rightarrow\ket{4}$ & 24 & 5.953 & 36 & 5.551 & 22\\
$\ket{5}\rightarrow\ket{6}$ & 24 & 5.953 & 32 & 5.462 & 26\\
$\ket{7}\rightarrow\ket{8}$ & 24 & 5.953 & 31 & 5.385 & 29\\

\end{tabular}
\caption{Simulation parameters used for the first $\hat{E}^{\dag}$ (top half) and second $\hat{E}^{\dag}$ (bottom half)}
\label{tab:params}
\end{table}
 
\begin{table}[h]
\centering
\begin{tabular}{c|c|c|c|c|c|c|c|c|c}
 $m$ & 0  & 1 & 2 & 3 & 4 & 5 & 6 & 7 & 8 \\ \hline
$\Phi_m$ & 0 & -1.589& -0.716& -0.907& 3.037& -1.621& 2.009& 2.859& -0.573 \\

\end{tabular}
\caption{Values of phases used for different Fock states $\ket{m}$ for SNAP gates}
\label{tab:SNAP}
\end{table}


\begin{thebibliography}{}
\bibitem{Footnote2}
A complementary method of amplifying SCSs (continuous-amplification) could be performed in specially engineered circuit-QED \cite{Yale_arxiv}.
\bibitem{Footnote01}
$\hat{E}^\dag$ is an isomorphism but is not unitary since $\hat{E}^\dag\hat{E}\ne \mathbb{I}$.
\end{thebibliography}
\end{document}